 \documentclass[final,5p,times,twocolumn,numbers]{elsarticle}

\usepackage{amssymb}
\usepackage{lipsum}

\usepackage{lineno}

\usepackage{hyperref}

\makeatletter
\def\Hy@appendixstring{Appendix}
\makeatother
\usepackage{amsmath}
\usepackage{subcaption}
\usepackage{makecell} 

\journal{Nuclear Physics A}

\begin{document}

\begin{frontmatter}



\title{Proton Energy Dependence of Radiation Induced Low Gain Avalanche Detector Degradation}


\author[label1,label2]{Veronika Kraus\corref{cor1}}
\ead{veronika.kraus@cern.ch}
\cortext[cor1]{Corresponding author}

\author[label3]{Marcos Fernández García}
\author[label1]{Luca Menzio} 
\author[label1]{Michael Moll}


\affiliation[label1]{organization={CERN, Organisation europénne pour la recherche nucléaire},
            addressline={Espl. des Particules 1},
            city={Genève},
            postcode={1217},
            country={Switzerland}}

\affiliation[label2]{organization={TU Wien, Faculty of Physics},
            addressline={Wiedner Hauptstraße 8-10},
            city={Vienna},
            postcode={1040},
            country={Austria}}

\affiliation[label3]{organization={Instituto de Física de Cantabria, IFCA (CSIC-UC)},
            addressline={Avda. los Castros},
            city={Santander},
            postcode={39005},
            country={Spain}}

\begin{abstract}
Low Gain Avalanche Detectors (LGADs) are key components for precise timing measurements in high-energy physics experiments, including the High Luminosity upgrades of the current LHC detectors. Their performance is, however, limited by radiation induced degradation of the gain layer, primarily driven by acceptor removal. This study presents a systematic comparison of how the degradation evolves with different incident proton energies, using LGADs from Hamamatsu Photonics (HPK) and The Institute of Microelectronics of Barcelona (IMB-CNM) irradiated with 18~MeV, 24~MeV, 400~MeV and 23~GeV protons and fluences up to $2.5\times10^{15}~\text{p/cm}^2$. Electrical characterization is used to extract the acceptor removal coefficients for different proton energies, whereas IR TCT measurements offer complementary insight into the gain evolution in LGADs after irradiation. Across all devices, lower energy protons induce stronger gain layer degradation, confirming expectations. However, 400~MeV protons consistently appear less damaging than both lower and higher energy protons, an unexpected deviation from a monotonic energy trend. Conversion of proton fluences to 1~MeV neutron-equivalent fluences reduces but does not eliminate these differences, indicating that the standard Non-Ionizing Energy Loss (NIEL) scaling does not fully account for the underlying defect formation mechanisms at different energies and requires revision when considering irradiation fields that contain a broader spectrum of particle types and energies.
\end{abstract}


\end{frontmatter}

\section{Introduction}
\label{introduction}
Low Gain Avalanche Detectors (LGADs) have become a key technology for precision timing in high-energy physics applications and beyond, enabling timing resolutions below 50\,ps for minimum ionizing particles and laying the foundations of the timing layers in the High-Luminosity LHC (HL-LHC) detector upgrades of ATLAS and CMS. Their performance relies on a $p^{+}$ gain layer in the $p$-type bulk that provides controlled charge multiplication through impact ionization. However, this gain layer is sensitive to radiation damage, which leads to acceptor removal (e.g., \cite{Moll2020}, \cite{Himmerlich2023}, \cite{Kramberger2015}), and ultimately the loss of gain and therefore good timing performance. Understanding and mitigating radiation induced degradation is therefore essential to ensure LGAD operability in harsh irradiation environments. \\
Research on irradiated LGADs has so far largely concentrated on a restricted set of particle types and energies. Radiation damage is then commonly quantified through the Non-Ionizing Energy Loss (NIEL) scaling, which is used as the reference model for relating different particle types and energies. Yet simulations indicate that different proton energies introduce qualitatively different defect structures \cite{Huhtinen2002}, with low energy protons producing predominantly point defects and higher energy protons generating increasingly clustered defects, suggesting that the NIEL framework alone may be insufficient to predict LGAD degradation over a wide range of hadron energies. As shown in previous studies \cite{Kramberger2015}, low energy hadrons such as protons lead to faster LGAD degradation than, for instance, neutrons of comparable fluence. These findings motivate a systematic comparison of LGAD degradation across a broad range of proton energies. A refined understanding of these energy dependent effects is essential for improving LGAD radiation hardness, and could be the first step of revising NIEL scaling to ensure reliable sensor performance in complex irradiation fields.

\begin{table*}[t]
\centering
\small
\renewcommand{\arraystretch}{1.2}
\begin{tabular}{@{}l l l l l@{}}
\hline
\textbf{Proton Energy} &
\textbf{DUTs} &
\textbf{Hardness Factor \cite{RD50ProtonNIEL}} &
\textbf{Fluences [$10^{14}\,\mathrm{p/cm^2}$]} &
\textbf{Rel.\ Error} \\
\hline
18 MeV  
& HPK W25 
& 3.00 
& 4,\ 8,\ 15,\ 25 
& 20\% \\

24 MeV  
& HPK W25, W36 
& 2.56 
& 0.1,\ 1,\ 4,\ 8,\ 15,\ 25 
& 10\% \\

400 MeV 
& \makecell[l]{HPK W25, W36;\\ IMB-CNM W1, W2, W4} 
& 0.83 
& 0.171,\ 2.75,\ 4.17,\ 8.94,\ 38.2 
& 10\% \\

23 GeV  
& \makecell[l]{HPK W25, W36;\\ IMB-CNM W1, W2, W4} 
& 0.62 
& 0.139,\ 5.57,\ 10.9,\ 22.9,\ 35.8 
& 7\% \\
\hline
\end{tabular}
\caption{Overview of proton irradiation campaigns, including proton energies, devices under test (DUTs), corresponding hardness factors used to convert fluences to 1~MeV neutron equivalents, applied proton fluences, and relative fluence uncertainties.}
\label{tab:irradiation_overview}
\end{table*}

\section{Materials and Methods}
\label{materials_methods}
\subsection{Devices Under Test}
\label{dut}
The present study includes a wide variety of devices under test (DUTs) from the two different manufacturers Hamamatsu Photonics (HPK) and The Institute of Microelectronics of Barcelona (IMB-CNM). HPK DUTs from the prototype 2 wafers W25 and W36, similar to what will be employed in the CMS Endcap Timing Layer (ETL) or the ATLAS  High-Granularity Timing Detector (HGTD), are used. The two wafers slightly differ in terms of the gain layer as described in more detail in \cite{Curras2023IEEE}. The single devices feature an active area of $1.3 \times 1.3~\text{mm}^2$ and an active thickness of 50\,\text{\textmu m}, with a gain layer depletion voltage $V_{gl}$ in the range of 50--60\,V and a breakdown voltage $V_{break}$ around 140\,V for W25 and 200\,V for W36. \\
From CNM, three wafers with different levels of carbon enrichment added to the boron dopants in the gain layer from production run 15973 were included: wafer~W1 without carbon implantation, wafer~W2 with a carbon dose of $1\times 10^{14}~\text{cm}^{-2}$, and wafer~W4 with a dose of $3\times10^{14}~\text{cm}^{-2}$. Carbon is often introduced to improve the radiation tolerance of the gain layer by altering the processes of defect generation. The CNM devices feature the same active area of $1.3 \times 1.3~\text{mm}^2$ and an active thickness of 50~\text{\textmu m}, with $V_{gl}$ in the range of 30--50\,V and $V_{break}$ varying between 160\,V and 220\,V for all three wafers.
Since all samples from this CNM run lack an opening in the topside metallization, laser characterization cannot be performed with this subset of devices.

\subsection{Proton Irradiation}
Proton irradiation was carried out at four different facilities: 18~MeV proton irradiation at the Bern University Hospital, 24~MeV irradiation at the University of Birmingham, 400~MeV irradiation at Fermilab and 23~GeV irradiation at the CERN PS-IRRAD facility. The respective subsets of samples included in each of the four irradiation campaigns are listed in \autoref{tab:irradiation_overview}. The five targeted proton fluences requested from the facilities were kept roughly similar (as far as possible given the technical constraints of each facility) and lie between $1.0\times10^{13}$ and $2.5\times10^{15}~\text{p/cm}^2$. Due to technical issues, the lowest targeted fluence could not be delivered for the 18~MeV protons, while for the 24~MeV irradiation an additional fluence point at $1.0\times10^{14}~\text{p/cm}^2$ was included. The actual received fluences and errors are listed in more detail in \autoref{tab:irradiation_overview}. With the hardness factors also listed in \autoref{tab:irradiation_overview}, the proton fluences can be converted to 1~MeV neutron equivalent fluence ($\mathrm{n_{eq}}$). The reported hardness factors for the four proton energies are taken from simulations \cite{RD50ProtonNIEL}, but were also cross-checked experimentally by evaluating the leakage current increase of reference diodes irradiated to the same proton fluences in order to confirm their consistency.

\section{Electrical Characterization}
\label{electrical_characterization}

\begin{figure}[t]
    \centering
    \includegraphics[width=0.45\textwidth]{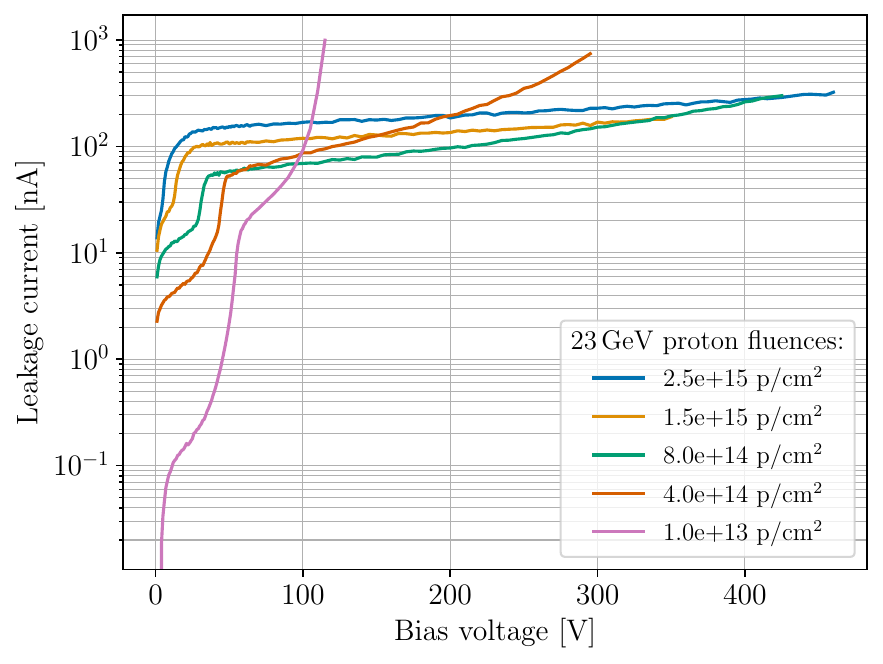}
    \caption{Exemplary I--V characteristics of HPK W25 LGADs after 23~GeV proton irradiation with increasing particle fluences. The steep rise in leakage current, attributed to the depletion of the gain layer, shifts toward lower applied bias voltages with increasing fluence due to degradation of the gain layer.}
    \label{fig:IV}
\end{figure}

\begin{figure*}[t]
    \centering
    \begin{subfigure}{0.45\textwidth}
        \centering
        \includegraphics[width=\textwidth]{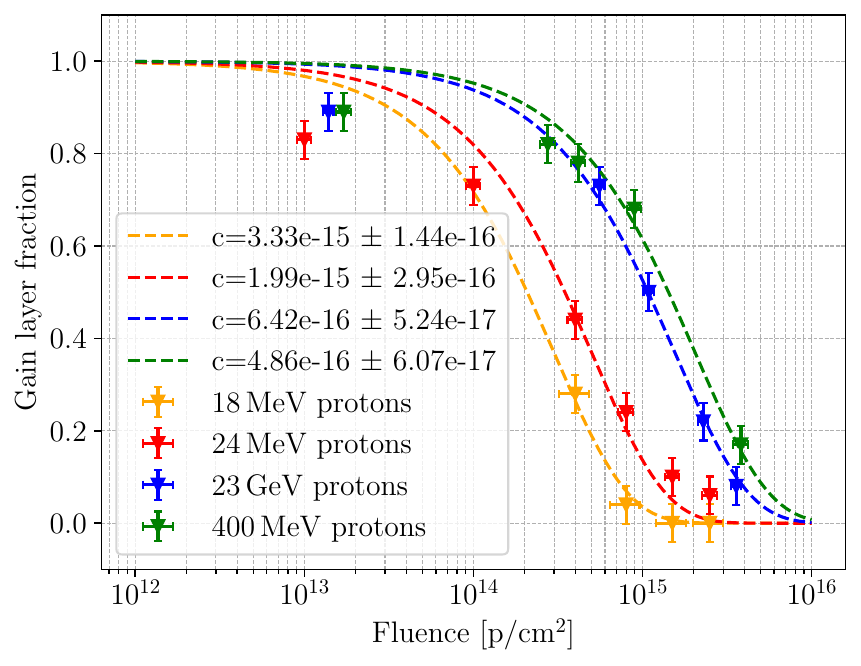}
        \caption{HPK LGADs Wafer 25}
        \label{subfig:HPK_W25}
    \end{subfigure}
    \hfill
    \begin{subfigure}{0.45\textwidth}
        \centering
        \includegraphics[width=\textwidth]{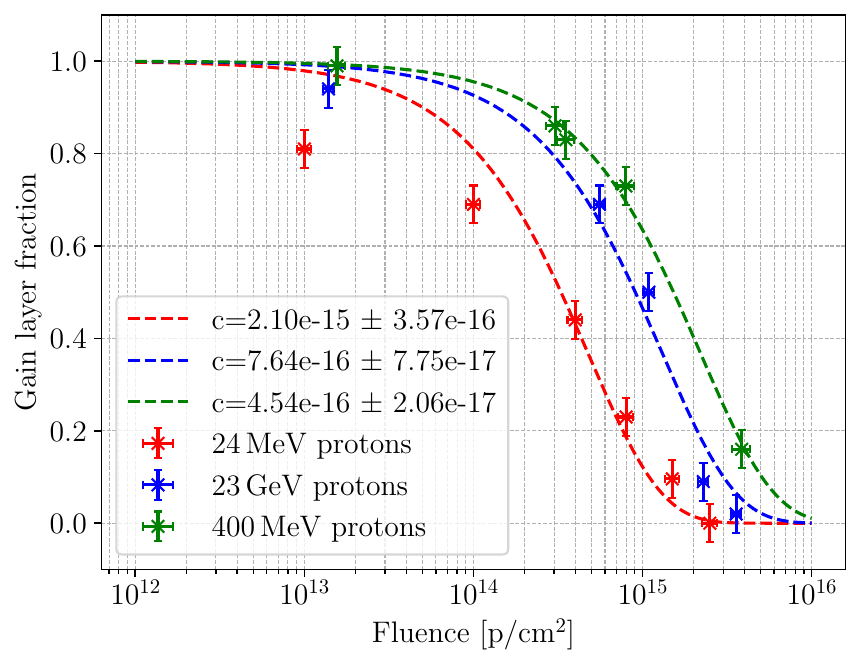} 
        \caption{HPK LGADs Wafer 36}
        \label{subfig:HPK_W36}
    \end{subfigure}
    \caption{The two plots show the degradation of HPK LGAD gain layers in terms of the reduction of $V_{gl}$ with increasing fluence for various proton energies. The fitted acceptor removal coefficients indicate a strong energy dependence of the damage, with 400~MeV protons consistently leading to the weakest degradation.}
    \label{fig:Acc_removal_HPK}
\end{figure*}

\begin{figure*}[t]
    \centering
    \begin{subfigure}{0.45\textwidth}
        \centering
        \includegraphics[width=\textwidth]{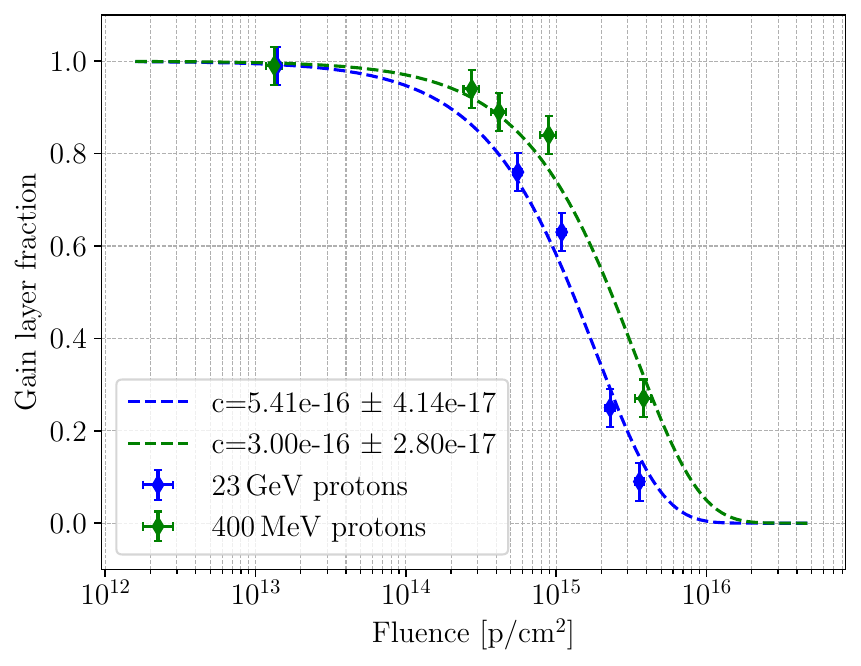}
        \caption{CNM LGADs Wafer 4}
        \label{subfig:CNM_E}
    \end{subfigure}
    \hfill
    \begin{subfigure}{0.45\textwidth}
        \centering
        \includegraphics[width=\textwidth]{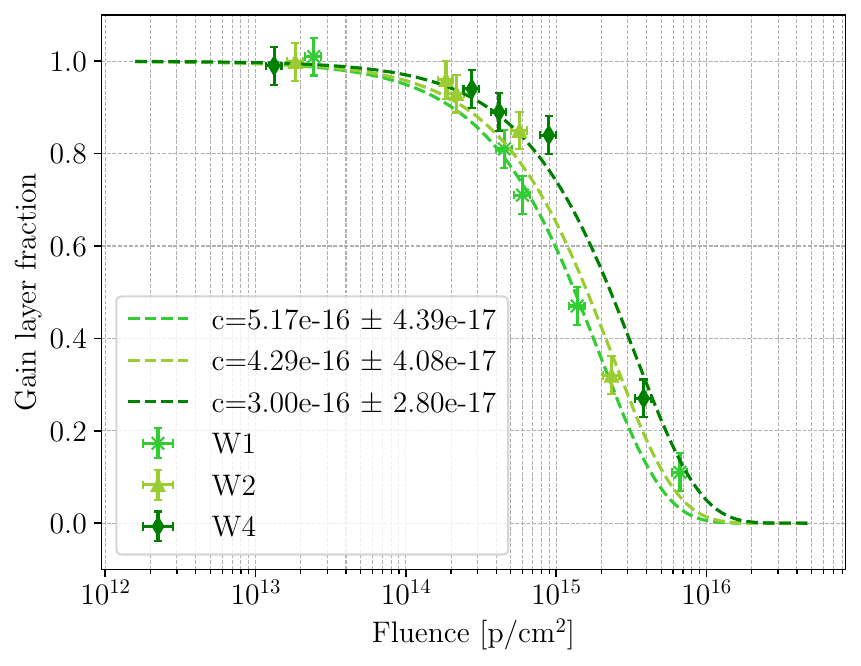} 
        \caption{CNM LGADs Wafter 1, 2, and 4}
        \label{subfig:CNM_W}
    \end{subfigure}
    \caption{The two plots show the acceptor removal behavior in CNM LGADs, comparing proton energies (left) and the three wafers with different carbon enrichment for one energy (right). The results illustrate both the weaker damage at 400~MeV compared to 23~GeV and the mitigation of acceptor removal with carbon in the gain layer.}
    \label{fig:Acc_removal_CNM}
\end{figure*}
\label{acceptor_removal_coefficients}

Electrical characterization was performed at the SSD laboratory probe station at CERN. The individual samples are contacted with two micro-positioning needles, one on the readout pad and the second on the guard ring, with the guard ring at ground potential. The high voltage bias is applied from the backside through the chuck. Current--voltage (I--V) and capacitance--voltage (C--V) scans were recorded before and after irradiation under identical conditions, except that the chuck is set to room temperature before and to $-20^{\circ}\mathrm{C}$ after irradiation of the devices. The characterization before irradiation gives $V_{gl}$ and $V_{break}$ for the different types of devices, as already mention in \autoref{dut}. After irradiation the same trends are observed for all proton energies and types of devices: Leakage current increases as expected due to defect creation, $V_{\mathrm{gl}}$ shifts towards lower applied bias voltages with increasing fluence as a consequence of gain layer degradation, and $V_{\mathrm{break}}$ shifts to higher bias voltages which can also be attributed to the degradation of gain layer and therefore loss of impact ionization. One set of exemplary I--V curves after 23~GeV proton irradiation with increasing fluences is shown in \autoref{fig:IV}. 

\subsection{Acceptor Removal Coefficients}

 \begin{figure}[t]
    \centering
    \includegraphics[width=0.45\textwidth]{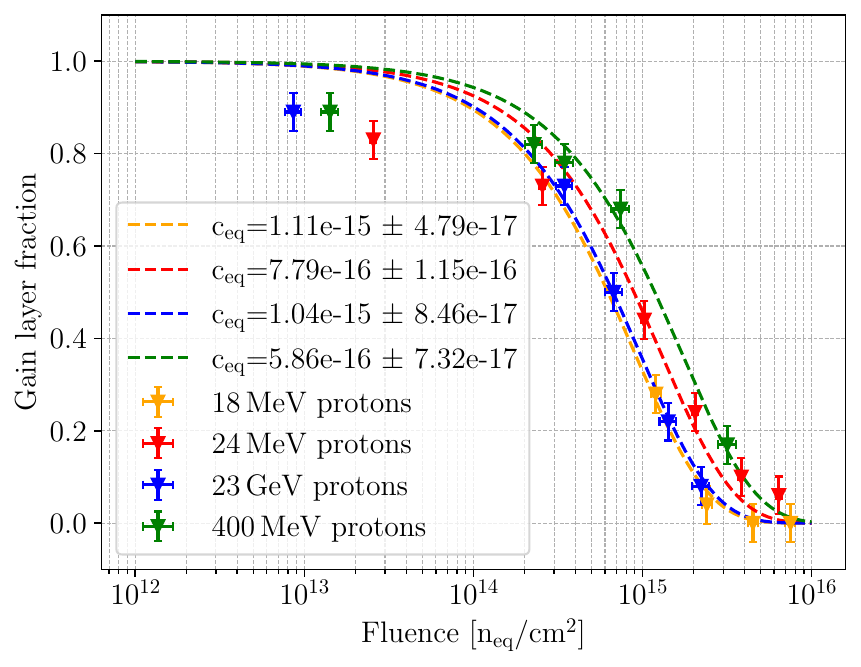}
    \caption{$V_{gl}$ of HPK2 W25 LGADs plotted versus $\mathrm{n_{eq}}$ fluence, showing still energy dependent acceptor removal behavior after the fluence conversion but with changed relative order compared to \autoref{subfig:HPK_W25}.}
    \label{fig:Acc_removal_HPK_neq}
\end{figure}

Based on the electrical characterization, $V_{gl}$ can be determined by locating the point of maximum slope in the I--V curve corresponding to gain layer depletion. It is obtained by extracting the maximum of 
\begin{equation}
k(I,V) = \frac{\Delta I}{\Delta V} \, \frac{V}{I} .
\end{equation}
After normalization to the non-irradiated reference, the fluence dependence of $V_{gl}$ is fitted to extract the acceptor removal coefficient from the relation
\begin{equation}
\frac{V_{gl}(\Phi)}{V_{gl}(0)} \approx e^{-c\,\Phi}
\quad \text{or} \quad
\frac{V_{gl}(\Phi_{\mathrm{eq}})}{V_{gl}(0)} \approx e^{-c_{\mathrm{eq}}\,\Phi_{\mathrm{eq}}}, 
\end{equation}
depending on whether the fluence is expressed as the proton fluence or as the neutron equivalent fluence n$_{\mathrm{eq}}$. The acceptor removal coefficient $c$ is used to compare the gain layer degradation caused by different proton energies. Plots of the evolution of $V_{gl}$ with increasing fluence and fitted $c$ value for different proton energies can be seen in \autoref{fig:Acc_removal_HPK} for HPK devices and \autoref{fig:Acc_removal_CNM} for CNM devices. \autoref{subfig:HPK_W25} shows acceptor removal for all four proton energies (18~MeV, 24~MeV, 400~MeV, 23~GeV) and devices of W25, while \autoref{subfig:HPK_W36} compares three proton energies for W36. The plots show a proton energy dependence of gain layer degradation: low energy protons (18--24~MeV) lead to the fastest degradation of the gain layer. 400~MeV protons consistently appear least damaging, yielding the smallest acceptor removal coefficients, also compared to the damage induced by the highest energy 23~GeV protons included in this study. The obtained $c$ value for HPK W25 LGADs irradiated with 23~GeV protons matches roughly results obtained in previous studies \cite{Curras2023}. For the CNM LGADs, the comparison between two proton energies for W4 in \autoref{subfig:CNM_E} shows the same trend of 400~MeV protons leading to less gain layer degradation than 23~GeV protons. \autoref{subfig:CNM_W} demonstrates the comparison for the same proton energy of 400~MeV between the three wafers with no (W1), moderate (W2) and high (W4) carbon enrichment. Carbon implantation mitigates acceptor removal, with higher carbon doses exhibiting systematically smaller $c$ values and thus a slower degradation. \\
The proton fluences for different energies can be converted into $\mathrm{n_{eq}}$ fluence using the corresponding hardness factor $\kappa$ given in \autoref{tab:irradiation_overview}, according to the relation 
\begin{equation}
\Phi_{\mathrm{eq}} = \kappa \cdot \Phi .
\end{equation}
For HPK W25 devices, the curves of $V_{gl}$ versus $\mathrm{n_{eq}}$ fluence for the different proton energies in \autoref{fig:Acc_removal_HPK_neq} appear to move closer together and partially change their relative ordering compared to the previously discussed \autoref{subfig:HPK_W25} with proton fluences on the $x$-axis. Still, the differences in the fitted $c_{\mathrm{eq}}$ values remain significant, demonstrating that NIEL scaling only partially compensates for the different damage mechanisms induced by different particle types and energies. After converting the fluences, the 400~MeV data still shows the weakest gain layer degradation.

\section{Laser Characterization}

\begin{figure*}[t]
    \centering
    \begin{subfigure}{0.45\textwidth}
        \centering
        \includegraphics[width=\textwidth]{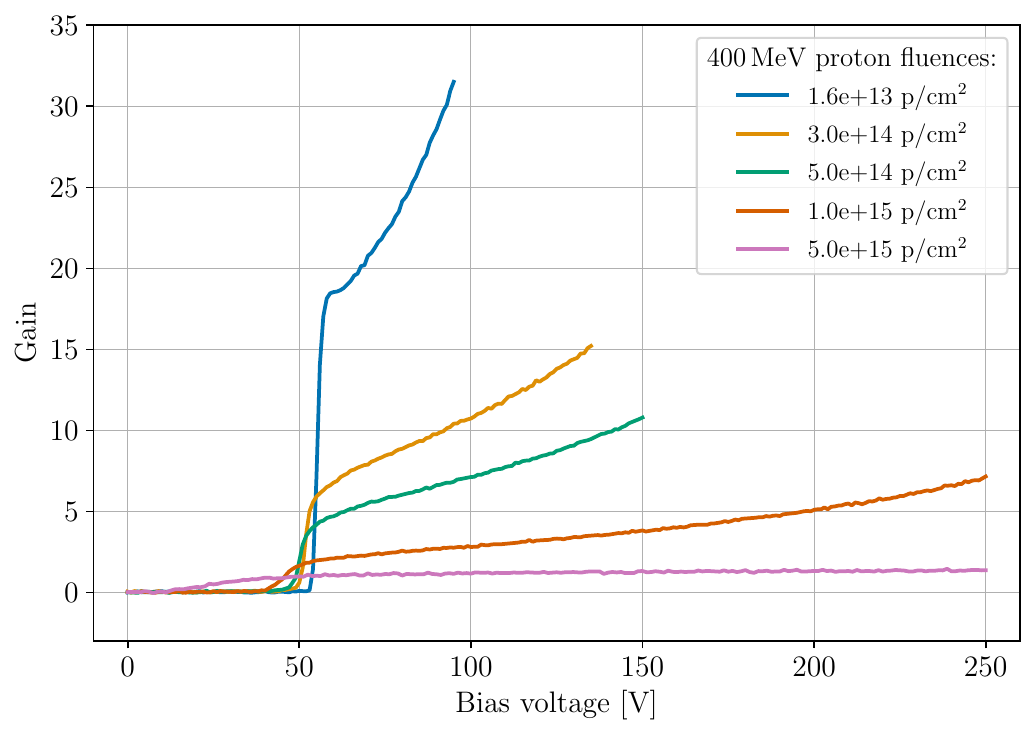}
        \caption{}
        \label{subfig:Gain_400MeV}
    \end{subfigure}
    \hfill
    \begin{subfigure}{0.45\textwidth}
        \centering
        \includegraphics[width=\textwidth]{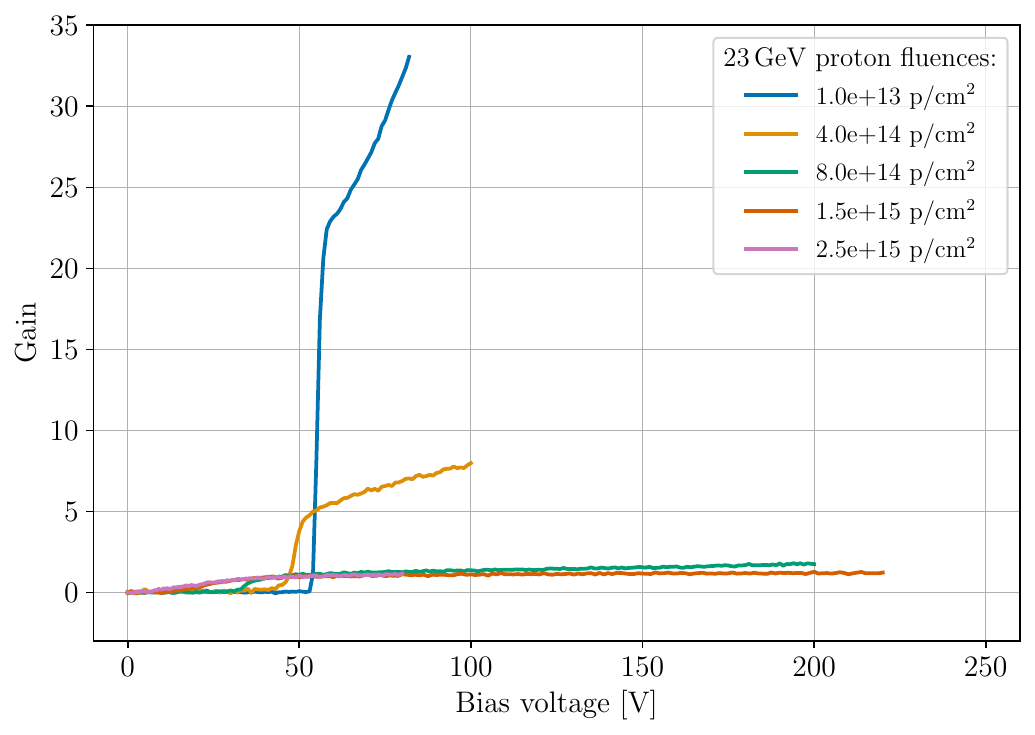} 
        \caption{}
        \label{subfig:Gain_23GeV}
    \end{subfigure}
    \caption{Gain measurements for HPK W25 LGADs irradiated with 400\,MeV and 23\,GeV protons show similar behavior at lower fluences. While 400\,MeV devices retain measurable gain up to about $10^{15}\,\mathrm{n_{eq}}$ fluence, faster loss of gain is observed for 23\,GeV irradiated devices at higher fluences.}
    \label{fig:Gain_comp}
\end{figure*}

\begin{figure}[t]
    \centering
    \includegraphics[width=0.45\textwidth]{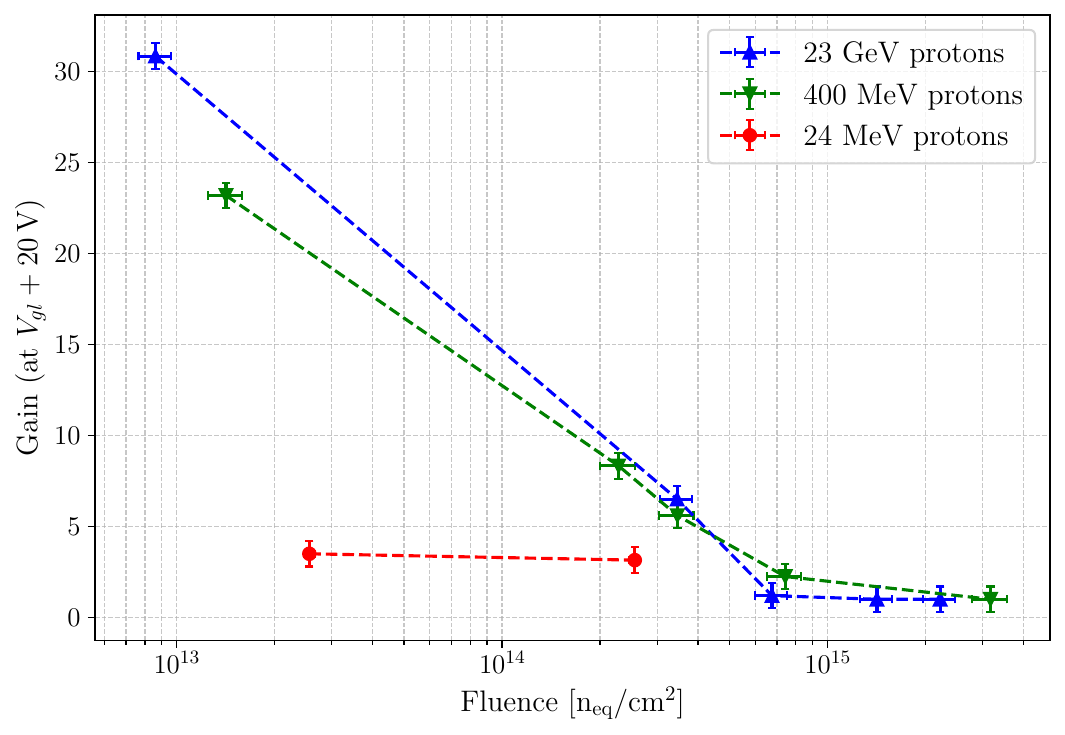}
    \caption{Gain at a fixed relative bias $(V_{gl} + 20~\mathrm{V})$ as a function of $\mathrm{n_{eq}}$ fluence for HPK W52 LGADs comparing irradiation with 24~MeV, 400~MeV, and 23~GeV protons. The data illustrate the strong energy dependence of gain degradation, with low energy protons causing the fastest loss of gain and 400~MeV protons leading to the slowest degradation at high fluences.}
    \label{fig:Gain_overview}
\end{figure}

While I--V and C--V measurements capture gain layer degradation due to acceptor removal, laser characterization provides complementary sensitivity to charge collection efficiency and trapping effects in the irradiated devices. Laser characterization was performed at the SSD Transient Current Technique (TCT) setup at CERN, using an IR laser with $\lambda=1064$\,nm and a spot size of $\approx 10$\,\text{\textmu m}. Since the CNM devices have no opening in the topside metallization, the TCT studies can only be performed for HPK devices, with a special focus on W25 since these devices where irradiated with all four proton energies. All laser characterization was carried out at a temperature of $-20^{\circ}\mathrm{C}$. The laser intensity settings for this study were kept equivalent to 10 minimum ionizing particles (MIPs). The samples are mounted on passive PCBs using conductive glue and wire bonds, allowing the IR laser pulses to be injected through the front side. The output is fed into a CIVIDEC broadband amplifier and subsequently digitized with a KEYSIGHT MXR254A oscilloscope. From the averaged, recorded waveforms, the collected charge ($CC$) can be determined through integration. Furthermore, the gain is calculated by comparing the $CC$ of the irradiated LGAD with the respective, fully depleted reference diode using 
\begin{equation}
\mathrm{Gain}[V] =
\frac{CC_{\mathrm{LAGD}}[V]}
     {CC_{\mathrm{PIN}}[V \ge V_{\mathrm{depl}}]} .
\end{equation}
Gain as a function of applied bias voltage for increasing particle fluences can be seen in \autoref{subfig:Gain_400MeV} for 400\,MeV and in \autoref{subfig:Gain_23GeV} for 23\,GeV proton irradiation. The gain--voltage curves show that both 400\,MeV and 23\,GeV proton irradiated LGADs exhibit similar behavior for the first two irradiation steps, with gain values around ten reached for fluences of $2$--$3\times10^{14}\,\mathrm{n_{eq}}$/cm$^2$. As the fluence increases, however, the 23\,GeV devices experience fast loss of gain above approximately $5\times10^{14}\,\mathrm{n_{eq}}$/cm$^2$. In contrast, the 400\,MeV irradiated samples show a more gradual degradation and retain measurable gain even up to about $10^{15}\,\mathrm{n_{eq}}$/cm$^2$. From the individual gain--voltage curves measured at different fluences and proton energies, the gain at a fixed relative bias $(V_{gl} + 20\,\mathrm{V})$ can be extracted and compared, as shown in \autoref{fig:Gain_overview}. This representation allows the gain degradation to be compared directly as a function of fluence across proton energies. The resulting trend shows that 23\,GeV and 400\,MeV protons lead to similar gain evolution, with 23\,GeV samples exhibiting slightly higher gain at low fluences, while 400\,MeV samples maintain measurable gain longer at high fluences. In contrast, the 24\,MeV protons cause a much faster reduction in gain, producing more damage in the devices. Overall, the qualitative behavior agrees with the findings from the electrical characterization: lower energy protons are more damaging to the gain layer, whereas 400\,MeV protons appear to be part of an energy regime producing particularly low damage even compared to the highest proton energy of 23\,GeV. 

\section{Summary and Outlook}
\label{summary_outlook}
This study presents an overview of how different proton energies affect LGAD degradation, combining electrical and laser characterization to extract acceptor removal coefficients and gain behavior over a wide fluence range. Despite the importance of these effects, there is limited literature directly comparing irradiation induced LGAD degradation as a function of proton energy, even though the underlying damage mechanisms are known to vary significantly with incident energy \cite{Huhtinen2002}. Low energy protons in particular highlight the limitations of the commonly used NIEL scaling, which does not, for example, distinguish between point defect and cluster defect formation in silicon. \\
In particular, acceptor removal coefficients $c$ are extracted for various types of HPK and CNM LGADs for four proton energies (18\,MeV, 24\,MeV, 400\,MeV, and 23\,GeV) from the evolution of $V_{gl}$ with fluence, enabling a quantitative comparison of the gain layer degradation due to acceptor removal. Laser characterization is carried out for the irradiated HPK W25 LGADs, with the gain determined by comparing their collected charge to that of reference diodes. Both characterization methods indicate that low energy protons (18--24~MeV) are most damaging, 23 GeV protons show moderate degradation, while 400~MeV protons consistently appear least damaging, a non-monotonic behavior that is not yet fully understood. This unexpected trend raises questions about the underlying defect formation mechanisms, suggesting that further simulations and modeling are needed to explain the energy dependence beyond standard NIEL scaling. \\
In addition to the measurements presented here, more detailed investigations are planned: For CNM samples, radioactive source measurements will be performed to complement the electrical characterization in devices without an opening in the topside metallization. Moreover, dedicated timing studies will be carried out for all devices to evaluate how the timing resolution evolves with proton energy and fluence. 

\section*{Acknowledgements}
This work has received support from the European Union’s Horizon Europe Research and Innovation Programme under Grant Agreement No. 101057511 (EURO-LABS).

\bibliographystyle{elsarticle-num} 


\end{document}